\documentclass[12pt,draftcls,journal,letterpaper,twosides,onecolumn]{IEEEtran}

\usepackage[]{chapterbib} 
\usepackage{amsmath,amssymb,amscd,latexsym,dsfont,wasysym,bbm,stmaryrd}
\usepackage{float}
\usepackage{color}
\usepackage{graphicx,subfigure,balance}
\usepackage{multirow,multicol}
\usepackage{verbatim}
\usepackage{psfrag}
\usepackage{comment}
\usepackage{makeidx}
\usepackage[]{algorithm}
\usepackage{algorithmic}
\usepackage{cite}
\usepackage{arydshln} 
\newcommand{\tr}[1]{\textrm{#1}}
\newcommand{\mr}[1]{\mathrm{#1}}
\newcommand{\tnr}[1]{{\textnormal{#1}}}

\newcommand{\mc}[1]{\mathcal{#1}}

\newcommand{\ms}[1]{\mathds{#1}}

\newcommand{\ov}[1]{\overline{#1}}

\newcommand{\bp}{\boldsymbol{p}}

\newcommand{\figref}[1]{Fig.~\ref{#1}}

\newcommand{\secref}[1]{Sec.~\ref{#1}}

\newcommand{\defref}[1]{Definition~\ref{#1}}
\newcommand{\exref}[1]{Example~\ref{#1}}

\newcommand{\propref}[1]{Proposition~\ref{#1}}

\newcommand{\ie}{i.e.,~} 		
\newcommand{\eg}{e.g.,~}	
\newcommand{\cf}{cf.~}		

\renewcommand{\emptyset}{\varnothing} 
\newcommand{\argmax}{\mathop{\mr{argmax}}}

\newcommand{\set}[1]{\{#1\}}

\newcommand{\cd}{\cdot}
\newcommand{\ld}{\ldots}

\newcommand{\pdf}{p}            			

\newcommand{\Ex}{\ms{E}}     			

\newcommand{\mcJ}{\mc{J}}

\newcommand{\mcL}{\mc{L}}

\newcommand{\mcP}{\mc{P}}

\newcommand{\msL}{\ms{L}}

\newcommand{\SNR}{\mathsf{snr}}  
\newcommand{\SNRrv}{\mathsf{SNR}}  
\newcommand{\SNRav}{\ov{\mathsf{snr}}}  
\newcommand{\SNRvec}{\boldsymbol{\mathsf{snr}}}  
\newcommand{\SNRrvvec}{\boldsymbol{\mathsf{SNR}}}  

\newcommand{\PF}{\tnr{PF}}  
\newcommand{\SC}{\tnr{SC}}  
\newcommand{\TS}{\tnr{TS}}  
\newcommand{\RR}{\tnr{RR}}  
\definecolor{refkey}{rgb}{0.2,0.8,0.6} 
\definecolor{labelkey}{rgb}{0.2,0.3,0.4} 

\usepackage[acronym,nonumberlist]{glossaries} 
\newacronym[\glsshortpluralkey=PDFs,\glslongpluralkey=probability density functions]{pdf}{PDF}{probability density function}
\newacronym[\glsshortpluralkey=CDFs,\glslongpluralkey=cumulative density functions]{cdf}{CDF}{cumulative density function}
\newacronym[\glsshortpluralkey=PMFs,\glslongpluralkey=probability mass functions]{pmf}{PMF}{probability mass function}
\newacronym[]{cm}{CM}{coded modulation}
\newacronym[]{pam}{PAM}{pulse amplitude modulation}
\newacronym[]{bpsk}{BPSK}{binary phase shift keying}
\newacronym[]{qam}{QAM}{quadrature amplitude modulation}
\newacronym[]{psk}{PSK}{phase shift keying}
\newacronym[\glsshortpluralkey=LLRs,\glslongpluralkey=logarithmic likelihood ratios]{llr}{LLR}{logarithmic likelihood ratio}
\newacronym[]{map}{MAP}{maximum a posteriori}
\newacronym[]{ml}{ML}{maximum likelihood}
\newacronym[]{md}{MD}{multiuser diversity}
\newacronym[\glsshortpluralkey=MIs,\glslongpluralkey=mutual informations]{mi}{MI}{mutual information}
\newacronym[\glsshortpluralkey=GMIs,\glslongpluralkey=generalized mutual informations]{gmi}{GMI}{generalized mutual information}
\newacronym[]{bicm-gmi}{BICM-GMI}{BICM generalized mutual information}
\newacronym[]{awgn}{AWGN}{additive white Gaussian noise}
\newacronym[]{amc}{AMC}{adaptive modulation and coding}
\newacronym[]{sp}{SP}{set-partitioning}
\newacronym[]{gsm}{GSM}{global system for mobile communications}
\newacronym[]{edge}{EDGE}{enhanced data rates for GSM evolution}
\newacronym[]{3gpp}{3GPP}{3rd generation partnership project}
\newacronym[]{dvb}{DVB}{digital video broadcasting}
\newacronym[\glsshortpluralkey=CCs,\glslongpluralkey=convolutional codes]{cc}{CC}{convolutional code}
\newacronym[\glsshortpluralkey=PCCCs,\glslongpluralkey=parallel concatenated convolutional codes]{pccc}{PCCC}{parallel concatenated convolutional code}
\newacronym[\glsshortpluralkey=TCs,\glslongpluralkey=turbo codes]{tc}{TC}{turbo code}
\newacronym{ldpc}{LDPC}{low-density parity-check}
\newacronym[]{ofdm}{OFDM}{orthogonal frequency-division multiplexing}
\newacronym[]{bep}{BEP}{bit-error probability}
\newacronym[]{blep}{BLEP}{block-error probability}
\newacronym[]{sep}{SEP}{symbol-error probability}
\newacronym[]{ttcm}{TTCM}{turbo-trellis coded modulation}
\newacronym[]{uep}{UEP}{unequal error protection}
\newacronym[\glsshortpluralkey=CENCs,\glslongpluralkey=convolutional encoders]{cenc}{CENC}{convolutional encoder}
\newacronym[]{mimo}{MIMO}{multiple-input multiple-output}
\newacronym[\glsshortpluralkey=SNRs,\glslongpluralkey=signal-to-noise ratios]{snr}{SNR}{signal-to-noise ratio}
\newacronym[]{msb}{MSB}{most significant bit}
\newacronym[]{bcjr}{BCJR}{Bahl--Cocke--Jelinek--Raviv}
\newacronym[\glsshortpluralkey=SEDs,\glslongpluralkey=squared Euclidean distances]{sed}{SED}{squared Euclidean distance}
\newacronym[\glsshortpluralkey=EDs,\glslongpluralkey=Euclidean distances]{ed}{ED}{Euclidean distance}
\newacronym[\glsshortpluralkey=MEDs,\glslongpluralkey=minimum Euclidean distances]{med}{MED}{minimum Euclidean distance}
\newacronym[]{core}{CoRe}{constellation rearrangement}
\newacronym[]{msd}{MSD}{multistage decoding}
\newacronym[]{pdl}{PDL}{parallel decoding of the individual levels}
\newacronym[\glsshortpluralkey=GCs,\glslongpluralkey=Gray codes]{gc}{GC}{Gray code}
\newacronym[]{brgc}{BRGC}{binary-reflected Gray code}
\newacronym[]{nbc}{NBC}{natural binary code}
\newacronym[]{fbc}{FBC}{folded-binary code}
\newacronym[]{bsgc}{BSGC}{binary semi-Gray code}
\newacronym[]{msp}{MSP}{modified set-partitioning}
\newacronym[]{ssp}{SSP}{semi set-partitioning}
\newacronym[]{fhd}{FHD}{free Hamming distance}
\newacronym[]{mfhd}{MFHD}{maximum free Hamming distance}
\newacronym[]{ods}{ODS}{optimal distance spectrum}
\newacronym[]{iud}{i.u.d.}{independent and uniformly distributed}
\newacronym[]{ud}{u.d.}{uniformly distributed}
\newacronym[]{iid}{i.i.d.}{independent and independently distributed}
\newacronym[]{bico}{BICO}{binary-input continuous-output}
\newacronym[]{gh}{GH}{Gauss--Hermite}
\newacronym[]{sl}{SL}{Shannon limit} 

\newacronym[]{lhs}{l.h.s.}{left-hand side}
\newacronym[]{rhs}{r.h.s.}{right-hand side} 
\newacronym[]{ra}{RA}{resource allocation}
\newacronym[]{rr}{RR}{round-robin}
\newacronym[]{er}{ER}{equal-rate}
\newacronym[]{pf}{PF}{proportional fairness}
\newacronym[]{ts}{TS}{time-sharing}
\newacronym[]{sc}{SC}{superposition coding}
\newacronym[]{pf-ts}{PF-TS}{proportionally fair TS}
\newacronym[]{pf-sc}{PF-SC}{proportionally fair SC}
\newacronym[]{er-ts}{ER-TS}{equal-rate TS}
\newacronym[]{er-sc}{ER-SC}{equal-rate SC}

\newacronym[\glsshortpluralkey=BSs,\glslongpluralkey=base-stations]{bs}{BS}{base-station}
\newacronym[\glsshortpluralkey=MSs,\glslongpluralkey=mobile-stations]{ms}{MS}{mobile-stations}
\newacronym[]{kkt}{KKT}{Karush--Kuhn--Tucker} 
\newacronym[]{mcs}{MCS}{modulation/coding scheme}

\IEEEoverridecommandlockouts

\title{Resource Allocation for Downlink Channel Transmission Based on Superposition Coding}
\author{Redouane Sassioui, Aata El Hamss, Leszek Szczecinksi,~\IEEEmembership{Senior Member,~IEEE},\\ and~Mustapha Benjillali,~\IEEEmembership{Member,~IEEE}
\\
\thanks{%
The work was supported by the government of Quebec, under grant \#PSR-SIIRI-435.}%
\thanks{%
R. Sassioui and M. Benjillali are with the Communications Systems Department, INPT, Rabat, Morocco. [e-mails: sassioui.redouane@gmail.com, benjillali@ieee.org]. R. Sassioui was also with INRS-EMT when the work was submitted for publication.}%
\thanks{%
A. El Hamss and L. Szczecinski are with INRS-EMT, Montreal, Canada. [e-mails: \{elhamss, leszek\}@emt.inrs.ca].}%
}%

\newtheorem{definition}{Definition}
\newtheorem{proposition}{Proposition}

\newtheorem{example}{Example}

\begin{document}

\maketitle

\begin{abstract}
We analyze the problem of transmitting information to multiple users over a shared wireless channel. The problem of \gls{ra} for the users with the knowledge of their channel state information has been treated extensively in the literature where various approaches trading off the users' throughput and fairness were proposed. The emphasis was mostly on the \gls{ts} approach, where the resource allocated to the user is equivalent to its time share of the channel access.  In this work, we propose to take advantage of the broadcast nature of the channel and we adopt \gls{sc}---known to outperform \gls{ts} in multiple users broadcasting scenarios. In \gls{sc}, users' messages are simultaneously transmitted by superposing their codewords with different power fractions under a total power constraint. The main challenge is to find a simple way to allocate these power fractions to all users taking into account the fairness/throughput tradeoff. We present an algorithm with this purpose and we apply it in the case of popular \gls{pf}. The obtained results using \gls{sc} are illustrated with various numerical examples where, comparing to \gls{ts}, a rate increase between 20\% and 300\% is observed.
\end{abstract}


\section{Introduction}\label{Sec:Introduction}
In this paper, we derive an \gls{ra} scheme for downlink multi-user communications where various utility functions may be applied. The distinctive feature of the analyzed scheme is that it is based on \gls{sc}. Unlike the popular and well studied \gls{ts} approach, where at each time instant only one user is receiving data, with \gls{sc} many users may receive their respective payload simultaneously.

In downlink communications over time-varying channels, \gls{ra} depends on the instantaneous channel condition between the \gls{bs} and the user (or \gls{ms}). This usually results in transmission schemes which allocate resources (time, frequency, power) to the user which experiences the most favourable channel conditions.

In presence of multiple users, it was shown in \cite{Knopp95} that the optimal strategy to maximize the total throughput (sum-rate of all users) is to schedule the user with the best link during each transmission unit. This \gls{md} \cite[Ch.~6.6]{Tse05_Book} maximizes the overall system throughput by allocating the shared resource to the user that can best exploit it.
However, this approach raises a ``fairness'' issue since it would result in shared resources being monopolized by  the users with the best channel conditions (\eg with a direct link to the \gls{bs}, or at a short range from it), while the user with poor channel conditions would rarely access the channel affecting considerably his throughput. %

Total throughput enhancement and fairness are hence crucial but conflicting criteria in the design of optimal \gls{ra} schemes.

To address this issue, many utility-based approaches---where utility represents a function of user's throughput---have been proposed in the literature to consider both fairness and throughput in the design of scheduling and \gls{ra} algorithms. Among them, \gls{pf} \cite{Kelly97}  based on the logarithmic utility function is a well-known criterion introduced to balance between throughput and fairness. Other approaches adopt different variants of the utility function but most of them can be reduced to the maximization of the weighted sum of users' throughputs. 

Using \gls{pf} (or any other utility-based criterion) in the case of \gls{ts} leads to well-known and simple-to-implement results with the channel being allocated to a single user at any transmission time \cite{Kelly97,Berggren04}. On the other hand, it is also known that \gls{ts} approach is outperformed by \gls{sc} \cite[Ch.~15.1.3]{Cover06_Book} when communicating over shared (broadcast) channels. In \gls{sc}, the transmitter splits the available power among the multiple users, \emph{superimposes} the resulting codewords, and broadcasts them on the downlink channel. The underlying assumption is that the users are capable of decoding \gls{sc} signals via successive decoding. This is not a very restrictive assumption as the so-called hierarchical modulation, closely related to \gls{sc}, is nowadays included in communication standards, \eg \cite{Chari07}.

\gls{sc}-based \gls{ra} for cooperative communications was analyzed in \cite{Kaneko11,Thai13} but the formal analysis of multi-user \gls{sc} was not addressed therein. It was also studied in  \cite{Shaqfeh09,Zafar13}, where optimal solutions were derived using the approach of \cite{Tse97,Tse97b}. With respect to \cite{Tse97,Shaqfeh09,Zafar13} our contributions are the following:
\begin{itemize}
\item We derive the power-fraction allocation algorithm from the \gls{kkt} conditions applied directly to the \gls{ra} problem at hand which is similar in spirit to the approach used by \cite{Zhang08} for the case of \gls{ts}. The resulting, sorting-like algorithm is very simple and has the complexity linear in the number of users (hundreds of users are easily dealt with). Our approach does not require the utility-based formalism of \cite{Tse97,Tse97b}; it is hence simpler to derive and reveals the underlying structure of relationships, which lead  to the simple algorithm we propose.
\item In the numerical examples we show that the gains provided by \gls{sc} combined with \gls{pf} criterion can lead to a multi-fold throughput increase for certain classes of users without penalty to the others. Moreover, we show that, in a single-cell scenario, and with a growing number of users, the total throughput improves by up to 50\%  with respect to \gls{ts}.
\item We observe that almost the entire power is distributed amongst just a few users. Motivated by this observation, we propose to apply \gls{sc} only to a limited number of users and propose the respective algorithms in this case.
\item We show that the two-user \gls{sc} is not only much more practical than the general multi-level superposition, but it also achieves most of the gains provided by unconstrained \gls{ra}. 
\end{itemize}

The rest of the paper is organized as follows. In \secref{Sec:Model}, we introduce the adopted transmission model, and in \secref{Sec:Sch.MD} we discuss \gls{ra} principles. We develop a simple algorithm to define the power allocation policy for \gls{sc} in \secref{Sec:PF.SC}, where we also analyze the case of power allocation under constraints on the number of scheduled users. 
Conclusions are drawn in \secref{Sec:Conclusions}.
\section{Transmission Model}\label{Sec:Model}
We consider the scenario where the \gls{bs} has to send information to $L$ distinct users. We consider the flat block-fading channel model commonly used in the analysis of wireless systems. Namely, we assume that at each discrete time instant $n$, the signal received by the $l$th user is modeled as 
\begin{align}\label{}
y_{l}[n]=\sqrt{\SNR_{l}[n]}x[n]+z_{l}[n], \quad l=1,\ld, L,
\end{align}
where $x[n]$ is the unitary-power signal emitted by the \gls{bs}, $z_{l}[n]$ is the zero-mean  unitary variance random process modeling noise/interference, and $\SNR_{l}[n]$ is the \gls{snr} at the $l$-th receiver. 

In the block-fading model, for a given user $l$, the \gls{snr} is modelled as a white random process $\SNRrv_{l}[n]$. Thus, the \gls{snr} remains constant for the duration of the entire block but varies independently between blocks. 

While we do not need to assume any particular distribution to characterize the fading, we focus on the Rayleigh distribution in the numerical examples, that is, the \gls{pdf} of  $\SNRrv$ is given by
\begin{align}\label{}
\pdf_{\SNRrv_{l}}(\SNR) = \frac{1}{\SNRav_{l}}\exp\!\left(-\frac{\SNR}{\SNRav_{l}}\right),
\end{align}
where $\SNRav_{l}$ is the average \gls{snr} of the $l$-th link.

The data of each user is assumed available at the \gls{bs} at any time instant $n$ (the so-called ``saturation'' scenario) and it is delay-insensitive, thus we can consider long-term averages as relevant performance measures.  Moreover, we assume that at the beginning of each transmission block, each user informs the \gls{bs} about the value of its instantaneous \gls{snr} $\SNR_{l}[n]$, through a perfect feedback channel. We do not consider the related transmission overhead here as this issue is out of the scope of the paper.

These assumptions allow us to focus on the main problem addressed in this work; namely, multi-user resource allocation, and in particular -- the one based on \gls{sc}.

The \gls{bs} at time instant $n$ forms the signal $x[n]$ using the \gls{mcs} $\phi(\cd)$ so that the rate conveyed to the user $l$ is given by
\begin{align}\label{rl.def}
r_{l}[n]=\phi_{l}(\SNRvec[n],\bp[n]),
\end{align}
where $\bp[n]$ gathers all the parameters defining the \gls{mcs} and 
\begin{align}\label{snr.vec}
\SNRvec[n]=[\SNR_{1}[n],\ld,\SNR_{L}[n]].
\end{align}

\gls{ra} consists, therefore, in choosing the appropriate vector $\bp[n]$. 

The simple and popular multi-user \gls{mcs} relies on time-sharing (\gls{ts}), where each user is assigned a fraction of the available transmission time so that
\begin{align}\label{phi.AWGN}
\phi_{l}^{\TS}(\SNRvec[n],\bp[n])=p_{l}[n]\log(1+\SNR_{l}[n]),
\end{align}
where, for simplicity, we assume that \gls{mcs} uses a capacity-achieving coding. That is, we consider the case, where signals $x_{l}$ are obtained from infinite-length, randomly generated Gaussian codebook. These idealistic assumptions allow us to focus on the allocation strategies and provide upper limits on the rates achievable for any practical coding scheme.

In the context of \gls{ts}, \gls{ra} consists most often in dedicating the entire transmission time to one particular user. Then, \emph{scheduling} (\ie determining which user should transmit) is equivalent to a \gls{ra}. The simplest \gls{ra} scheme is based on the so-called \gls{rr} approach where each user is assigned periodically (with period $L$) to the entire transmission block, thus
\begin{align}\label{}
\bp[n]&=\boldsymbol{\delta}_{t^{\tr{RR}}[n]}\\
t^{\tr{RR}}[n]&= \llbracket n \rrbracket _{L} +1
\end{align}
where we use $\boldsymbol{\delta}_{t}=[0, \ld, 0, 1, 0, \ld, 0]$ to denote the $L$-length vector with a non-zero element at position $t$, and $\llbracket \cd\rrbracket _{L}$ denotes  the modulo-$L$ operation.

Then, each user occupies the channel during exactly the same fraction $1/L$ of the overall transmission time and its throughput is given by 
\begin{align}\label{RR.R.l}
R_{l}=\frac{1}{L}\Ex_{\SNRrv_{l}}[ \log(1+\SNRrv_{l}) ].
\end{align}
We note that the same result in terms of throughput will be obtained assigning each user a portion $p_{l}=1/L$ of the  block (if we ignore the practical limitation related to distributing the finite time among $L$ users).

\section{Resource Allocation}\label{Sec:Sch.MD}

\gls{ra} strategies may be defined via a function $\bp=\bp(\SNRvec)$ designed to maximize the sum of the so-called utility functions defined over users' throughputs $R_{l}$
\begin{align}\label{sum.U}
\hat{\bp}(\SNRvec)=\argmax_{\bp(\SNRvec)} \sum_{l=1}^{L} U(R_{l}),
\end{align}
where
\begin{align}\label{Ex.R.l}
R_{l}=\Ex_{\SNRrvvec}\left[\phi_{l}\bigl(\SNRrvvec,\bp(\SNRrvvec)\bigr)\right]
\end{align}
and
\begin{align}\label{}
\SNRrvvec=[\SNRrv_{1},\ld,\SNRrv_{L}]
\end{align}
is the random vector modeling \eqref{snr.vec}.

For example, using $U(R)=R$ corresponds to the maximization of the aggregate throughput $R=\sum_{l=1}^{L}R_{l}$ and it can be shown that, then, the optimal \gls{ra} is defined via \gls{ts} with only one user (having the maximum instantaneous rate (MR) or --equivalently, the maximum \gls{snr}) scheduled for transmission within the block \cite{Berggren04}, \ie
\begin{align}
\label{max.SNR.1}
\bp[n]&=\boldsymbol{\delta}_{t^{\tr{MR}}[n]}\\
\label{max.SNR.2}
t^{\tr{MR}}[n]&=\argmax_{l\in\set{1,\ld,L}}\SNR_{l}[n].
\end{align}

However, \gls{ra} in \eqref{max.SNR.1}-\eqref{max.SNR.2} results in a situation where the high-\gls{snr} users receive the highest throughput $R_{l}$, while weak-\gls{snr} users obtain lower throughputs $R_{l}$. This is considered ``unfair'' \cite{Berggren04}.

To address this issue, various criteria have been proposed in the literature aiming to improve the fairness of \gls{ra} algorithms. Among them,  the \gls{pf} criterion is arguably one of the most popular \cite{Kelly97}\cite{Zhang08} and corresponds to \eqref{sum.U} based on the utility function 
\begin{align}\label{Ux.log}
U(R)=\log(R).
\end{align}

On the other hand, the max-min optimization
\begin{align}\label{}
\hat{\bp}=\argmax_{\bp}\min_{l\in\set{1,\ld,L}}\set{R_{l}}
\end{align}
where resources are allocated so that the weakest user is prioritized, tend to yield \gls{er} \gls{ra}.

\subsection{On-line adaptation}

Using $U(R)=R$, the function $\bp(\SNRvec)$ is defined in closed-form via \eqref{max.SNR.1} and \eqref{max.SNR.2} but this is rarely the case. In fact, it is rather difficult to find the optimal mapping $\hat{\bp}(\SNRvec)$ for the popular utility function, such as the one in \eqref{Ux.log} corresponding to \gls{pf}. The main difficulty is to calculate the expectation \eqref{Ex.R.l} in closed form.

To overcome this problem, we may use estimates of the throughput based on temporal averages \cite{Zhang08}
\begin{align}
\tilde{R}_{l}[n]&=\frac{1}{W}\sum_{t=0}^{W-1}r_{l}[n-t]\nonumber\\
\label{R.n}
&=\frac{r_{l}[n]-r_{l}[n-W]}{W}+\tilde{R}_{l}[n-1].
\end{align}

Using \eqref{R.n} in \eqref{sum.U}, finding the optimal allocation parameters for the \gls{pf} utility function \eqref{Ux.log} can be formulated as the following optimization problem \cite{Zhang08}
\begin{align}
\label{sum.U.n.1}
\hat{\bp}[n]&=\argmax_{\bp[n]}\sum_{l=1}^{L} U\left(\tilde{R}_{l}[n-1]+\frac{r_{l}[n]-r_{l}[n-W]}{W}\right).
\end{align}
Further, for long observation windows, $W\rightarrow\infty$, \ie when 
\begin{align}\label{}
\frac{r_{l}[n]}{W}\rightarrow 0,
\end{align}
we may use the first-order approximation $U(\tilde{R}+r)\approx U(\tilde{R})+U'(\tilde{R})\cd r$, which yields
\begin{align}
\label{sum.U.n.3}
\hat{\bp}[n]&\approx\argmax_{\bp[n]}\sum_{l=1}^{L}U'\bigl(\tilde{R}_{l}[n-1]\bigr) r_{l}[n]\\
\label{sum.U.n.5}
&=\argmax_{\bp[n]}\sum_{l=1}^{L}\beta_{l}[n] r_{l}[n]
\end{align}
where the terms independent of $\bp$ and the common multiplication factor $W$ ( not affecting the optimization results) were removed.

The form of \eqref{sum.U.n.5}, emphasizes that the utility-function based approach may be reduced to the optimization of the sum of instantaneous rates $r_{l}[n]$ weighted by $\beta_{l}[n]=U'(\tilde{R}_{l}[n-1])$ \cite{Zafar13}. 

We emphasize that the adaptation rule \eqref{sum.U.n.5} is valid irrespectively of the adopted utility function or \gls{mcs}, that is, it may be applied for various forms of $\phi_{l}(\SNRvec,\bp)$ or $U(R)$. In particular, for $U(R)=R$ we recover the max-SNR (\ie also max-instantaneous rate $r_{l}[n]$) solution we have shown in \eqref{max.SNR.1}-\eqref{max.SNR.2}.

\begin{example}[Resource allocation in \gls{ts}]\label{ex:pf-ts}
Considering \gls{ts} again, we have to use \gls{mcs} with rates defined by \eqref{phi.AWGN}, thus
\eqref{sum.U.n.5} becomes
\begin{align}\label{pf.ts.Opt}
\hat{\bp}^{\PF-\TS}[n]&=\argmax_{\bp}\sum_{l=1}^{L} p_{l}\beta_{l}[n]\log_{2}(1+\SNR_{l}[n])\\
&\tnr{s.t.}\quad \sum_{l=1}^{L}p_{l}=1, p_{l}\geq 0.
\end{align}

It is easy to see that \eqref{pf.ts.Opt} is solved by scheduling only one user \cite{Zhang08} indexed by
\begin{align}\label{eq:pf.ts}
t[n]=\argmax_{l\in\set{1,\ld, L}}\beta_{l}[n]\log_{2}(1+\SNR_{l}[n]).
\end{align}

Then, if we opt for using the \gls{pf} utility function \eqref{Ux.log}, we obtain $U'(R)=R^{-1}$, thus $\beta_{l}[n]=1/\tilde{R}_{l}[n-1]$ and \eqref{eq:pf.ts} becomes
\begin{align}\label{eq:pf.ts.2}
t^{\PF-\TS}[n]=\argmax_{l\in\set{1,\ld, L}}\frac{\log_{2}(1+\SNR_{l}[n])}{\tilde{R}_{l}[n-1]}.
\end{align}
Thus, the optimal solution is given by 
\begin{align}\label{eq:pf.bp}
\hat{\bp}^{\PF-\TS}[n]=\boldsymbol{\delta}_{t^{\PF}[n]}.
\end{align}
\end{example}

This is the well-known \gls{pf-ts} resource allocation \cite{Liew08}. The choice of the scheduled user depends on the ratio (proportion) between the instantaneous achievable rate $\log(1+\SNR_{l}[n])$ and the throughput $\tilde{R}_{l}[n-1] \approx R_{l}$.  Thanks to the normalization by $R_{l}$, the users with relatively small average \gls{snr} (and thus also relative small value of $R_{l}$) are granted access to the channel more frequently than in the non-proportional max-\gls{snr} scheduling \eqref{max.SNR.1}-\eqref{max.SNR.2}.

We note that we do not  calculate explicitly the expectation \eqref{Ex.R.l}. Instead, by applying \eqref{eq:pf.ts} and \eqref{eq:pf.bp}, the \gls{ra} algorithm ``learns'' through the local optimization \eqref{pf.ts.Opt} what the globally optimal solution is.

An important common feature of all mentioned \gls{ra} schemes based on \gls{ts} is that, in the $n$th block, only one user is scheduled for transmission, that is, \eqref{eq:pf.ts} is valid independently of the chosen utility-function. 

\section{Optimal \gls{ra} with superposition coding}\label{Sec:PF.SC}

We will now take the analysis of \gls{ra} based on utility-function to a more involved multi-user \gls{mcs} well suited for the wireless downlink transmission. While we use the \gls{pf} utility in the examples, \ie $\beta_{l}[n]=1/\tilde{R}_{l}[n-1]$, the presented solutions will be general, and remain valid when the utility function changes.

To motivate the adoption of \gls{sc}, and before defining the \gls{ra} framework, we outline the principle of encoding/decoding based on \gls{sc}.

\subsection{\gls{sc} Broadcasting Principles}\label{Sec:2U.SC}
From an information-theoretic point of view, sending information to the users over a shared channel (\ie where the users receive the same broadcasted signal) is done optimally via \gls{sc} \cite[Ch.~15.1.3]{Cover06_Book}. 

In the case of $L=2$ users, the solution that maximizes the sum of weighted rates is obtained by transmitting a superposition of  the codewords, that is
\begin{align}\label{}
x[n]=\sqrt{p_{1}}x_{1}[n]+\sqrt{p_{2}}x_{2}[n]
\end{align}
where $x_{1}[n]$ and $x_{2}[n]$ are the unitary power signals of each user, $p_{1}$ and $p_{2}$ are their power fractions, and we impose the constraint $p_{1}+p_{2}=1$ so the  emitted signal $x[n]$ has a unitary power.

We assume without loss of generality that $\SNR_{1}\leq \SNR_{2}$. The decoding can be performaned as follows: the weak-\gls{snr} user decodes only its own message $x_{1}[n]$ (treating the signal $x_{2}[n]$ as interference). Since it receives the signal
\begin{align}\label{}
y_{1}[n]=\sqrt{\SNR_{1}p_{1}}x_{1}[n] + \sqrt{\SNR_{1}p_{2}}x_{2}[n] +z_{1}[n],
\end{align}
its achievable rate is given by
\begin{align}\label{rate.1.L2}
\phi_{1}^{\SC}(\SNRvec,\bp)=\log_{2}\left(1+\frac{p_{1}\SNR_{1}}{p_{2}\SNR_{1}+1}\right),
\end{align}
where the denominator of the fraction under the logarithm amalgams the power of the noise $z_{1}[n]$ as well as the interference created by the signal  $\sqrt{p_{2}}x_{2}[n]$, which is possible because both are independent Gaussian variables. 

User $l=2$ (with $\SNR_{2}\geq \SNR_{1}$) can also decode message $x_{1}[n]$ and remove it from the received signal 
\begin{align}\label{}
y_{2}[n]=\sqrt{\SNR_{2}}x[n]+z_{2}[n],
\end{align}
the decoding of his own message $x_{2}[n]$ relies then on the interference-free signal
\begin{align}\label{}
y'_{2}[n]=y_{2}[n]-\sqrt{\SNR_{2}[n]p_{1}}x_{1}[n]=\sqrt{\SNR_{2}[n]p_{2}}x_{2}[n]+z_{2}[n],
\end{align}
thus, the resulting rate is 
\begin{align}\label{rate.2.L2}
\phi_{2}^{\SC}(\SNRvec,\bp)=\log_{2}\left(1+p_{2}\SNR_{2}\right).
\end{align}

Since user $l=2$ discards the message contained in $x_{1}[n]$, decoding $x_{1}[n]$ does not contribute to his throughput.

Parameters $\bp=[p_{1},p_{2}]$ determine the power allocated to users and the level of interference user $l=1$ experiences due to the signal $x_{2}[n]$ of user $l=2$. We emphasize here that the powers are allocated to both users using solely instantaneous values of \glspl{snr}. That is, we do not attempt to take advantage of the channel dynamics, done by the so-called \emph{water-filling} algorithms.

\begin{figure}[htb]
\newcommand{\scale}{0.8}
\psfrag{xlabel}[c][c]{$\phi_{2}(\SNRvec,\bp)$}
\psfrag{ylabel}[c][c]{}
\psfrag{XX1-XXX}[cl][cl][0.9]{$\phi_{1}^{\SC}(\SNRvec,\bp)$}
\psfrag{X2}[cl][cl][0.9]{$\phi_{1}^{\TS}(\SNRvec,\bp)$}
\psfrag{p2=0}[bl][bl][\scale]{$p_{2}=0$}
\psfrag{p2=1}[bl][bl][\scale]{$p_{2}=1$}
\psfrag{R1}[bc][cc][\scale][90]{$\log(1+\SNR_{1})$}
\psfrag{Zx}[rc][rc][\scale]{$0$}
\psfrag{R2}[ct][ct][\scale]{$\log(1+\SNR_{2})$}
\psfrag{R2b}[ct][ct][\scale]{$\log(1+\SNR'_{2})$}
\psfrag{Zy}[ct][ct][\scale]{$0$}
\begin{center}
\scalebox{1}{\includegraphics[width=1\linewidth]{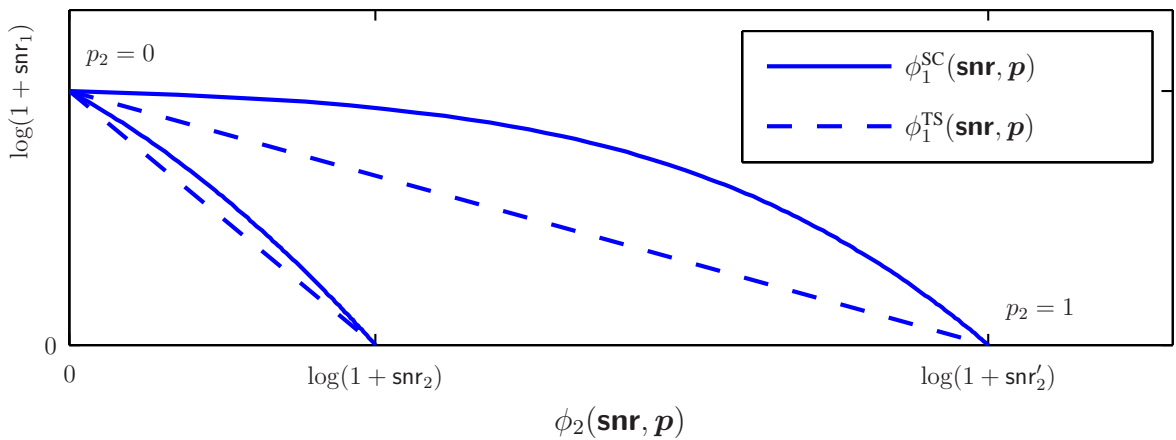}}
\caption{Rates $\phi_{1}(\SNRvec,\bp)$ vs. $\phi_{2}(\SNRvec,\bp)$ for $\bp=[1-p_{2}, p_{2}]$ achievable with \gls{sc} and \gls{ts} and $\SNR_{1}<\SNR_{2}<\SNR'_{2}$. All rates on the corresponding curves are achievable varying $p_{2}\in(0,1)$. }\label{Fig:SCvsTS}
\end{center}
\end{figure}

To get an insight into the potential gains, \figref{Fig:SCvsTS} compares the rates achievable with \gls{sc} and \gls{ts}. 

All pairs of transmission rates $\bigl(\phi_{1}(\SNRvec,\bp)$, $\phi_{2}(\SNRvec,\bp)\bigr)$  on the curves corresponding to \gls{ts} and \gls{sc}, can be obtained varying $p_{2}\in(0,1)$. Of course, the interpretation of the parameter $p_{2}$ depend on the \gls{mcs}: for \gls{ts} $p_{2}$ has a meaning of a time fraction, while for \gls{sc} it represents a fraction of the transmit power. Clearly, for any given rate $\phi_{1}(\SNRvec,\bp)$, using \gls{sc}, the rate of the remaining user $\phi_{2}(\SNRvec,\bp)$ can be always greater when comparing to the rate obtained via \gls{ts}. Exception are  $p_{2}=0$ (only user $l=1$ transmits) and $p_{2}=1$ (only user $l=2$ transmits), when \gls{ts} and \gls{sc} are equivalent. 

Moreover, we note that the advantage of using \gls{sc} instead of \gls{ts} becomes important when the difference between the \glspl{snr} of both users increases (note the difference between the \gls{sc} and \gls{ts} curves for $\SNR'_{2}>\SNR_{2}$); in fact, for $\SNR_{1}=\SNR_{2}$, \gls{ts} and \gls{sc} are equivalent.

\gls{sc} transmission can be generalized to the case of $L>2$ as follows: the transmitted signal is given by
\begin{align}\label{}
x[n]=\sum_{l=1}^{L}\sqrt{p_{l}}x_{l}[n]
\end{align}
where $\sum_{l=1}^{L}p_{l}=1$. 

Assuming $\SNR_{1}\leq \SNR_{2}\leq\ld\leq \SNR_{L}$, the decoding by the user $l$ is done similarly to the case of $L=2$: the signals of weak-SNR users are decoded and subtracted in a successive-interference-cancellation approach\footnote{When $L=2$, this is done only for $l=2$.}, while the signals of strong-SNR users are treated as interference. Then, the rate of a reliable transmission to user $l$ is given by
\begin{align}\label{r.k.L}
\phi_{l}^{\SC}(\SNRvec,\bp)=\log_{2}\left(1+\frac{p_{l}\SNR_{l}}{\ov{p}_{l}\SNR_{l}+1}\right),
\end{align}
where 
\begin{align}\label{ov.p}
\ov{p}_{l}=\sum_{j=l+1}^{L}p_{j}
\end{align} 
denotes the total power of users $l{+}1,l{+}2, \ld, L$. For convenience of notation, in what remains, we use
\begin{align}\label{sum.0}
\sum_{k=l}^{l-1}a_{k}\triangleq 0;
\end{align}
here, it means that $\ov{p}_{L}= 0$.

\subsection{Resource allocation for $L=2$}\label{Sec:2U.SC.RA}
We now consider the case of \gls{ra} for $L=2$, which is relatively simple to derive and reveals the more general relationships that will be used for arbitrary $L$.

If $\SNR_{1}<\SNR_{2}$, using \eqref{rate.1.L2} and \eqref{rate.2.L2} in \eqref{sum.U.n.5} we have to solve the following maximization problem
\begin{align}\label{p2.opt}
\hat{p}_{2}=\argmax_{p_{2}\in(0,1)}\log\frac{(1+p_{2}\SNR_{2})^{\beta_{2}}}{(1+p_{2}\SNR_{1})^{\beta_{1}}},
\end{align}
where, the constraint $p_{1}+p_{2}=1$ is taken into account and---to alleviate the notation---we omit time indices, \ie $\SNR_{k}\equiv \SNR_{k}[n]$ and $\beta_{k}\equiv \beta_{k}[n]$.

Similarly, for $\SNR_{2}<\SNR_{1}$, we need to solve
\begin{align}\label{p1.opt}
\hat{p}_{1}=\argmax_{p_{1}\in(0,1)}\log\frac{(1+p_{1}\SNR_{1})^{\beta_{1}}}{(1+p_{1}\SNR_{2})^{\beta_{2}}}.
\end{align}

After a simple algebra, the solution of \eqref{p2.opt} is given by:
\begin{align}\label{p2.opt.L2}
\hat{p}_{2}=
\begin{cases}
1, & \!\!\!\tr{if~~}  \beta_{1}\!\le\!\beta_{2} \vee 
\displaystyle{\frac{\beta_{1}\SNR_{1}}{1+\SNR_{1}}\!\le\!\frac{\beta_{2}\SNR_{2}}{1+\SNR_{2}}}\\
0, & \!\!\!\tr{if~~}  \beta_{1}\SNR_{1}\!\ge\!\beta_{2}\SNR_{2}\\
\displaystyle{\frac{\beta_{2}\SNR_{2}-\beta_{1}\SNR_{1}}{\SNR_{1}\SNR_{2}(\beta_{1} -\beta_{2} )}},& \!\!\!\tr{otherwise}.
\end{cases}
\end{align}

\begin{figure}[h]
\newcommand{\scale}{0.8}
\psfrag{xlabel}[ct][ct]{$\SNR_{1}$}
\psfrag{ylabel}[cb][cb]{$\SNR_{2}$}
\psfrag{k}[ct][ct][\scale]{{\small$\displaystyle{\frac{\beta_{1}}{\beta_{1}{-}\beta_{2}}}$}}
\psfrag{XX-1}[cc][cc][\scale]{$p_{2}=1$}
\psfrag{XX-2}[cc][cc][\scale]{$\hat{p}_{2}=0$}
\psfrag{XX-3}[cc][cc][\scale]{$\hat{p}_{2}=p_{2,1}$}
\psfrag{XX-4}[cc][cc][\scale]{$\hat{p}_{2}=0$}
\psfrag{XX-1}[cc][cc][\scale]{$\hat{p}_{2}=1$}
\psfrag{TS-Limit}[tc][bc][\scale][57]{$\beta_{1}\log_{2}(1+\SNR_{1})=\beta_{2}\log_{2}(1+\SNR_{2})$}
\psfrag{SNR=Limit}[tc][bc][\scale][35]{$\SNR_{1}=\SNR_{2}$}
\psfrag{SNRbeta=Limit}[bc][tc][\scale][42]{$\SNR_{1}\beta_{1}=\SNR_{2}\beta_{2}$}
\psfrag{tau=Limit}[tc][bc][1][85]{$\frac{\beta_{1}\SNR_{1}}{1+\SNR_{1}} = \frac{\beta_{2}\SNR_{2}}{1+\SNR_{2}}$}
\begin{center}
\scalebox{1}{\includegraphics[width=1\linewidth]{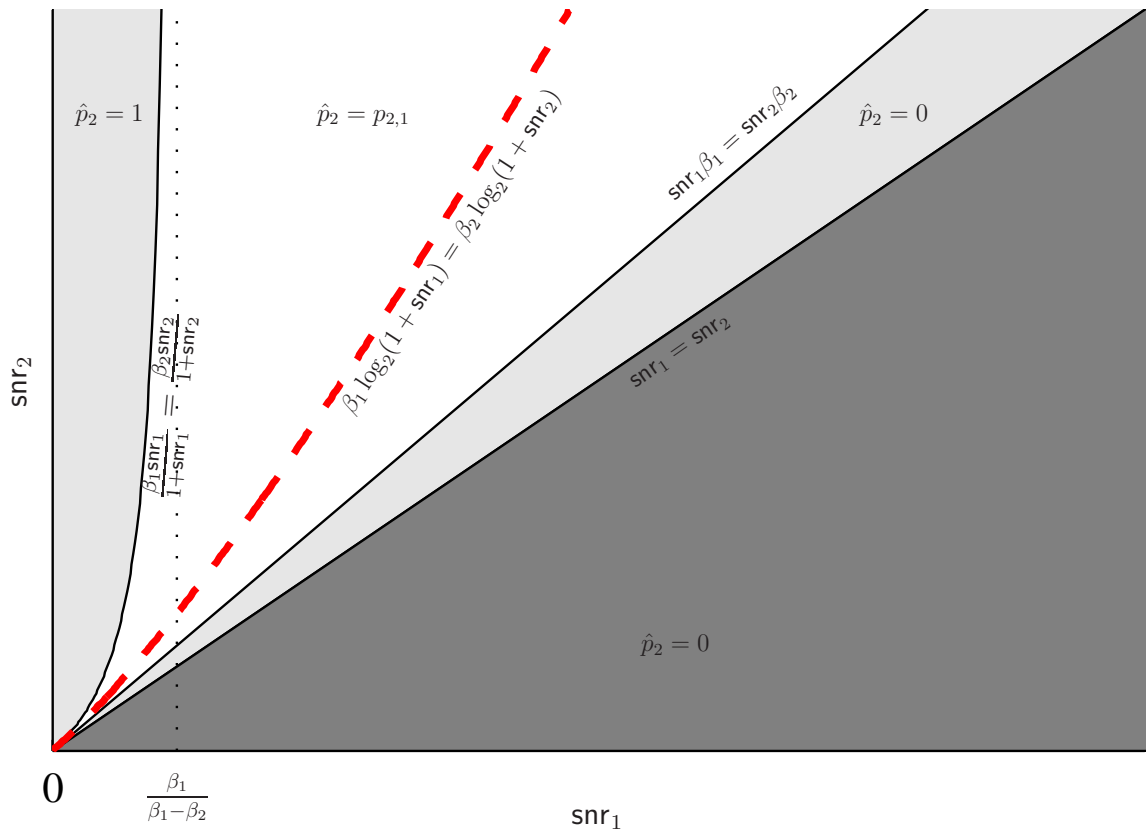}}
\caption{Resource allocation may be seen as a mapping $\SNRvec\rightarrow\hat{\bp}$. Here, $L=2$ and  $\beta_{1}>\beta_{2}$ and \gls{ra} is defined via \eqref{p2.opt.L2}. The light-shaded regions correspond to the solution of the problem \eqref{p2.opt} (solved under assumption $\SNR_{2}>\SNR_{1}$) where only one user is scheduled for transmission ($\hat{p}_{1}=1$ or $\hat{p}_{2}=1$). The unshaded region corresponds to the case where we use \gls{sc}, \ie $\hat{p}_{1},\hat{p}_{2}\in(0,1)$. In the  dark-shaded region we have $\SNR_{2}<\SNR_{1}$ so the solution is found solving the problem \eqref{p1.opt}: through symmetry to the first case of \eqref{p2.opt.L2},  if $\SNR_{2}<\SNR_{1}$ and $\beta_{1}>\beta_{2}$ we set $\hat{p}_{1}=1$. The thick dashed (red) line separates the decision regions of \gls{ts}: above the line we schedule the user $l=2$ (thus $\hat{p}_{2}=1$), while below the line we schedule user $l=1$ ($\hat{p}_{1}=1$).}\label{Fig:Regions.L2}
\end{center}
\end{figure}

Fixing $\boldsymbol{\beta}=[\beta_{1},\beta_{2}]$, the solution $\hat{\bp}$ depends solely on the values of the \glspl{snr}. In \figref{Fig:Regions.L2}, we illustrate how $\SNRvec$ affects the choice of $\hat{p}_{2}$ and compare \gls{sc} with \gls{ts}. Depending on the values of $\SNRvec$ and $\boldsymbol{\beta}$ we may obtain the solution equivalent to \gls{ts} (where we transmit to only one user) or to \gls{sc} where we transmit to both users simultaneously. 

\subsection{Arbitrary number of users}\label{L.Users}

Our objective function in \eqref{sum.U.n.5} is now defined as 
\begin{align}\label{y.SC}
y_{\msL}(\bp)=\sum_{l=1}^{L} \beta_{l}\phi^{\SC}_{l}(\SNRvec,\bp).
\end{align}
where  we use $\msL=\set{1,2\ld, L}$ --the set integers from 1 to $L$, to emphasize that all the users are considered as the candidates for \gls{ra}; later, in \secref{sec:RA.Kmax}, we will consider the optimization under restrictions on the users that may be scheduled.

Then we have to solve the following optimization problem
\begin{align}\label{Opt.SC}
\hat{\bp}=\argmax_{\bp} y_{\msL}(\bp), \quad \tnr{s.t.}\quad \sum_{l=1}^{L}p_{l}=1, \quad p_{l }\geq 0.
\end{align}

Applying the \gls{kkt} conditions, as done also in\cite{Liew08}, 
we know that there exists a Lagrange multiplier $\lambda$ (associated with the constraint $\sum_{l=1}^{L}p_{l}=1$) and multipliers $\mu_{l}, l=1,\ld, L$ (each, associated with the constraint $p_{l}\geq 0$) such that the optimal solution of \eqref{Opt.SC} satisfies 
\begin{align}\label{}
\label{KKT.1}
\frac{\partial y_{\msL}(\hat{\bp}) }{\partial p_{l}}  -\lambda +\mu_{l}=0,
\end{align}
where for brevity, we use $\displaystyle{\frac{\partial y_{\msL}(\hat{\bp}) }{\partial p_{l}}\triangleq\frac{\partial y_{\msL}(\bp) }{\partial p_{l}} |_{\bp=\hat{\bp}}}$. In \eqref{KKT.1} $\mu_{l}\geq0$; if $\mu_{l}>0$ we say that the positivity constraint $p_{l}\geq 0 $ is active, and then $\hat{p}_{l}=0$. If $\mu_{l}=0$ the constraint is inactive.  

Our problem will be then solved in two interconnected steps:
\begin{enumerate}
\item First, we find indices $l$ of the users which are not scheduled for transmission, that is, for which the positivity constraints are active (where we can thus set $\hat{p}_{l}=0$).
\item Next, we show that the remaining users have inactive positivity constraints and we explain how to calculate  their optimal power fractions $\hat{p}_{l}$.
\end{enumerate}

Using \eqref{KKT.1} we can conclude with respect to the parameters $p_{l}$ for which the positivity constraints are active. Namely, for any $j,k\in\msL$ we have
\begin{align}\label{SC_KKT_1_1}
 \frac{\partial y_{\msL}(\hat{\bp})}{\partial p_{j}}>\frac{\partial y_{\msL}(\hat{\bp})}{\partial p_{k}} \Rightarrow \hat{p}_{k}=0.
\end{align}

Assuming without any loss of generality that $j>k$, and after simple algebra, the \gls{lhs} of \eqref{SC_KKT_1_1} can be expressed as follows:
\begin{align}\label{SC_KKT_2}
\frac{\partial y_{\msL}(\hat{\bp})}{\partial p_{j}}>\frac{\partial y_{\msL}(\hat{\bp})}{\partial p_{k}} \Leftrightarrow \frac{\SNR_{j}\beta_{j}}{\SNR_{k}\beta_{k}}+\sum_{l=k+1}^{j-1}\hat{p}_{l}v_{l}>\frac{1+\hat{\ov{p}}_{k}\SNR_{j}}{1+\hat{\ov{p}}_{k}\SNR_{k}},
\end{align}
where $\sum_{l=k}^{k-1}a_{l}\triangleq 0$ takes care of the case $j=k+1$, and $v_{l}$ are arbitrary real numbers.

We want to establish conditions under which the inequality \eqref{SC_KKT_2} is satisfied irrespectively of $\hat{\ov{p}}_{k}$, which will allow us to identify the elements of $\hat{\bp}$ to be made equal to zero, \ie $\hat{p}_{k}=0$ or $\hat{p}_{j}=0$.

\begin{proposition}\label{prop1}
If $j>k$ and $\sum_{l=k+1}^{j-1}\hat{p}_{l}=0$, then the following relationships hold:
\begin{align}
\label{Prop1.1}
\beta_{k}&\leq\beta_{j} \vee \tau_{k}<\tau_{j}&  &\Rightarrow &\hat{p}_{k}=0,\\
\label{Prop1.3}
\beta_{k}&>\beta_{j} \wedge \nu_{k}>\nu_{j}&  &\Rightarrow &\hat{p}_{j}=0,
\end{align}
where
\begin{align}\label{}
\nu_{k}\triangleq\SNR_{k}\beta_{k},\\
\tau_{k}\triangleq\frac{\nu_{k}}{1+\SNR_{k}}
\end{align}
\begin{proof} \cf Appendix.
\end{proof}
\end{proposition}

\begin{definition}\label{def.1}
Denote by $p_{j,k}$ the solution of 
\begin{align}
\label{f.p.jk}
\frac{1+p\SNR_{j}}{1+p\SNR_{k}}=\frac{\nu_{j}}{\nu_{k}}
\end{align}
with respect to $p$.
\end{definition}


\begin{proposition}\label{prop2}
If $k<j<m$, $\sum_{l=k+1}^{j-1}\hat{p}_{l}=0$, and $\sum_{l=j+1}^{m-1}\hat{p}_{l}=0$, then the following holds:
\begin{align}\label{prop2.1}
0&\leq p_{j,k}\leq p_{m,j}\leq 1 & & \Rightarrow & \hat{p}_{j}=0.
\end{align}
\begin{proof} \cf Appendix.
\end{proof}
\end{proposition}

\propref{prop1} and \propref{prop2} allow us to ``purge'' users whose power fractions are zero $\hat{p}_{l}=0$. This can be done via simple element-by-element comparison between the parameters $\beta_{k}, \nu_{k}, \tau_{k}$, and $p_{j,k}$ using the algorithms we define below.  

To simplify the description of the algorithm, it is convenient to define a set $\mcL=\set{\ell_{1},\ell_{2},\ld,\ell_{K}}$ as the ordered set that gathers indices to $K$ ``non purged'' users, \ie for which we did not determine if $\hat{p}_{\ell_{k}}= 0 , k=1,\ld, K$. Then, purging user $l\in\mcL$ is equivalent to the elimination of his index $l$ from the set $\mcL$, which we denote as $\mcL\gets \mcL \backslash l$.

We start with Algorithm~\ref{Algo.1} which eliminates users according to \eqref{Prop1.1}. After this first purge, we use Algorithm~\ref{Algo.2} which enforces \eqref{Prop1.3}. Finally, we need to purge users using \propref{prop2} and, to this end, we proceed using  Algorithm~\ref{Algo.3}. 

\renewcommand{\algorithmicrequire}{\textbf{Input:}}
\renewcommand{\algorithmicensure}{\textbf{Output:}}
\begin{algorithm}[htb]
{\small
\caption{Purging users according to \eqref{Prop1.1}}	\label{Algo.1}
\begin{algorithmic}[1]
\REQUIRE $\beta_{l}, \tau_{l}$
\ENSURE Removes indices $l$ from the set $\mcL$ according to \eqref{Prop1.1}.
\STATE $\mcL \gets \msL$
\STATE $j\gets L$
\STATE $k\gets j-1$
\WHILE{$k\geq1$} 
	\IF{$\beta_{k}>\beta_{j} \wedge \tau_{k}>\tau_{j}$}
		\STATE $j \gets k$
	\ELSE
		\STATE $\mcL \gets \mcL \backslash k$
	\ENDIF
	\STATE $k\gets k-1$
\ENDWHILE
\end{algorithmic}
}
\end{algorithm}

\renewcommand{\algorithmicrequire}{\textbf{Input:}}
\renewcommand{\algorithmicensure}{\textbf{Output:}}
\begin{algorithm}[htb]
{\small
\caption{Purging users according to \eqref{Prop1.3}}	\label{Algo.2}
\begin{algorithmic}[1]
\REQUIRE $\nu_{l}, \mcL$
\ENSURE Removes indices $l$ from the set $\mcL$ according to \eqref{Prop1.3}.
\STATE $K \gets |\mcL|$
\STATE $k\gets 1$
\STATE $j\gets k+1$
\WHILE{$j\leq K$} 
	\IF{$\nu_{\ell_{k}}<\nu_{\ell_{j}}$}
		\STATE $k \gets j$
	\ELSE
		\STATE $\mcL \gets \mcL \backslash \ell_{j}$
	\ENDIF
	\STATE $j\gets j+1$
\ENDWHILE
\end{algorithmic}
}
\end{algorithm}

\renewcommand{\algorithmicrequire}{\textbf{Input:}}
\renewcommand{\algorithmicensure}{\textbf{Output:}}
\begin{algorithm}[htb]
{\small
\caption{Purging users according to \eqref{prop2.1} if $K>2$}	\label{Algo.3}
\begin{algorithmic}[1]
\REQUIRE $p_{k,j}, \mcL$
\ENSURE Removes indices $l$ from the set $\mcL$ according to \eqref{prop2.1}.
\STATE $K \gets |\mcL|$
\STATE $k\gets 1$
\STATE $j\gets k+1$
\STATE $m\gets k+2$
\WHILE{$m\leq K$} 
	\IF{$p_{\ell_{m},\ell_{j}}<p_{\ell_{j},\ell_{k}}$}
		\STATE $k \gets j$
	\ELSE
		\STATE $\mcL \gets \mcL \backslash \ell_{j}$
	\ENDIF
	\STATE $j \gets m$
	\STATE $m\gets m+1$
\ENDWHILE
\end{algorithmic}
}
\end{algorithm}

It is immediate to see that each of the above algorithms is executed using at most $L$ element-by-element comparisons. The total complexity is then linear in $L$.

After executing Algorithm~\ref{Algo.3}, $K=|\mcL|$ users remain unpurged. We can now determine the optimal power-fractions.

If $K=1$, \ie there is only one user with non-zero power fraction, \ie $p_{\ell_{1}}=1$.


\begin{proposition}\label{prop3} 
After applying Algorithm~\ref{Algo.1}, Algorithm~\ref{Algo.2}, and Algorithm~\ref{Algo.3}, the positivity constraints of all users remaining in the set $\mcL$ are inactive, \ie their Lagrange multipliers are $\mu_{\ell_{k}}=0, k=1,\ld, K$.
\begin{proof} \cf Appendix.
\end{proof}
\end{proposition}

Then, to find the power-fractions we can use the following.
\begin{proposition}\label{prop4}
If the number of users which are not purged via Algorithm~\ref{Algo.1}, Algorithm~\ref{Algo.2}, and Algorithm~\ref{Algo.3} is greater than one ($K>1$), the optimal solution of the problem in \eqref{Opt.SC} is found using the following rule:
\begin{align}
\label{prop3.1}
\hat{p}_{\ell_{K}}&=p_{\ell_{K},\ell_{K-1}}\\
\label{prop3.2}
\hat{p}_{\ell_{l}}&=p_{\ell_{l},\ell_{l-1}}-p_{\ell_{l+1},\ell_{l}}, \quad l=2,\ld, K-1, \\
\label{prop3.3}
\hat{p}_{\ell_{1}}&=1-p_{\ell_{2},\ell_{1}}
\end{align}	
\begin{proof} From \propref{prop3}, we know that $\mu_{\ell_{k}}=0$. Then, the optimality conditions in \eqref{KKT.1}, combined with \eqref{f.p.jk}, yield $p_{\ell_{j},\ell_{k}}=\hat{\ov{p}}_{\ell_{k}}$. This immediately yields the relationships in \eqref{prop3.1}, \eqref{prop3.2}, and \eqref{prop3.3}.
\end{proof}
\end{proposition}

\begin{example}[Optimal solution for $L=7$]

Suppose we have $L=7$ users with the following numerical values
\begin{align}\label{}
\SNRvec&=
[1.7,~ ~3.3,~ ~ ~4.4,~ ~6.7,~ ~7.7,~ ~8.3,~ ~8.6]\\
\boldsymbol{\beta}&=
[6.0,~ 29.7,~ 26.5,~ 15.4,~ 4.6,~ 17.6,~ 12.2]
\end{align}

We are thus able to calculate
\begin{align}\label{}
\boldsymbol{\nu}&=[10.2, 98.0, 116.6, 103.2, 35.4, 146.1,104.9]\\
\boldsymbol{\tau}&=[~3.8,~ 22.8, ~21.6,~ ~13.4,~ 4.1,~ 15.7,~ ~10.9].
\end{align}

Running Algorithm~\ref{Algo.1}, we obtain
\begin{align}
\label{}
\SNRvec&=[\times, ~3.3, ~~~4.4, \times, ~\!\times, ~ ~ ~\!8.3, ~~~8.6]\\
\label{nu.alog1}
\boldsymbol{\nu}&=[\times, 98.0, 116.6, \times, \times, 146.1, 104.9],
\end{align}
where we use ``$\times$'' to denote the irrelevant values corresponding to the purged users.

Using $\boldsymbol{\nu}$ from \eqref{nu.alog1} in Algorithm~\ref{Algo.2} we obtain
\begin{align}
\label{SNR.algo2}
\SNRvec&=[\times,~ 3.3, ~~4.4, \times, \times, ~8.3, \times]\\
\label{beta.algo2}
\boldsymbol{\beta}&=[\times, 29.7, 26.5, \times, \times, 17.6, \times].
\end{align}

The non-purged users are now indicated by the set \mbox{$\mcL=\set{2,3,6}$}, so we use \eqref{SNR.algo2} and \eqref{beta.algo2} in \eqref{f.p.jk} to calculate
\begin{align}\label{}
p_{6,3}=0.09, \qquad p_{3,2}=0.40,
\end{align}
and, after applying \propref{prop4}, we obtain the optimal solution
\begin{align}\label{}
\hat{p}_{6}&=0.09,\quad \hat{p}_{3}=0.31, \quad \hat{p}_{2}=0.60.
\end{align}

\end{example}

\begin{example}[Two groups of users and proportional fairness]\label{ex:AB.users}
We assume now that there are two groups of users, labeled ``A'' and ``B''. Each is composed, respectively, of $L_\tr{A}$ and $L_\tr{B}$ users having the same respective average \gls{snr}s, $\SNRav_\tr{A}$ and $\SNRav_\tr{B}$. In \figref{Fig:Rate.A.B}, we show the throughput per user in each group: $R_\tr{A}$ and $R_\tr{B}$, for \gls{ra} strategies based on \gls{rr}, \gls{pf-ts}, and \gls{pf-sc}. 

We make the following observations:

\begin{figure}[tb]
\psfrag{SNR2(GB)}[ct][ct]{$\SNRav_\tnr{B}$}
\psfrag{ylabel}[c][c]{BER}
\psfrag{Throughput}{}
\psfrag{GB-SC-XX}{$R_\tr{B}$, \SC} \psfrag{GA-SC}{$R_\tr{A}$, \SC}
\psfrag{GB-TS}{$R_\tr{B}$, \TS} \psfrag{GA-TS}{$R_\tr{A}$, \TS}
\psfrag{GB-RR}{$R_\tr{B}$, \RR} \psfrag{GA-RR}{$R_\tr{A}$, \RR}
\begin{center}
\scalebox{1}{\includegraphics[width=0.7\linewidth]{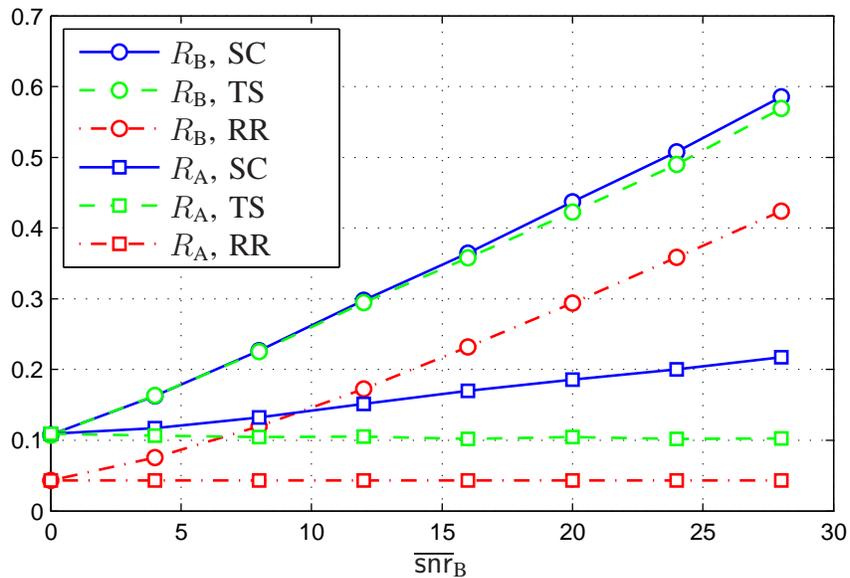}}\\
a)
\\
\scalebox{1}{\includegraphics[width=0.7\linewidth]{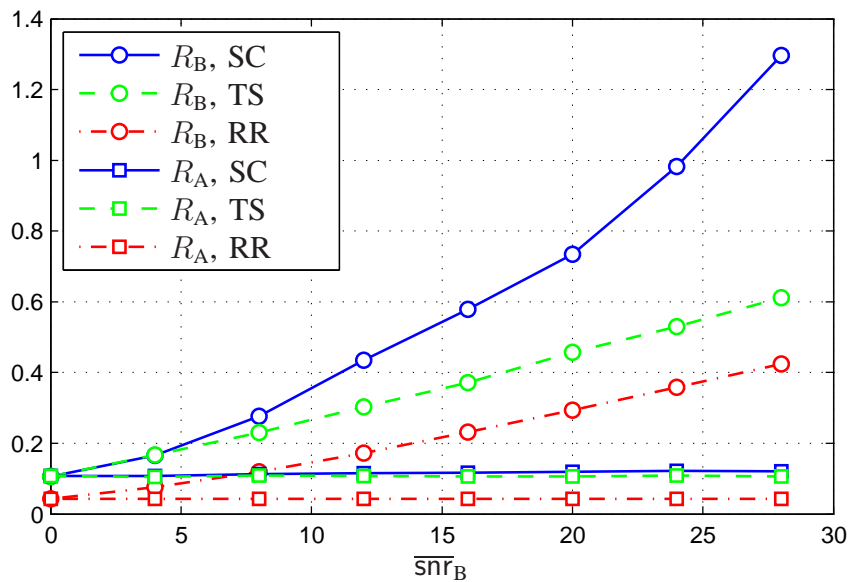}}\\
b)
\caption{The throughput obtained using \gls{rr}, \gls{pf-ts}, and \gls{pf-sc} resource allocation policies for a)~$L_\tr{A}=4$, $L_\tr{B}=16$ and b)~$L_\tr{A}=16$, $L_\tr{B}=4$; $\SNRav_\tr{A}=0$dB.}\label{Fig:Rate.A.B}
\end{center}
\end{figure}

\begin{enumerate}
\item The advantage of \gls{pf-sc} over \gls{pf-ts} is well pronounced when $\SNRav_\tr{A}$ and $\SNRav_\tr{B}$ differ significantly as then, \gls{sc} is most likely to provide notable gains. This is a reminiscence  of the broadcasting results for a fixed \gls{snr} shown in \secref{Sec:2U.SC}. 
\item Increasing the \gls{snr} of one group with respect to the other, the most significant throughput increase is obtained by the users in the least populated group irrespectively of their \gls{snr}:  their throughput grows by up to 100\% with respect to \gls{pf-ts}. For example, in \figref{Fig:Rate.A.B}a, we observe that increasing the \gls{snr} of group ``B'',  the throughput of users in group ``A'' improves by 100\%. This can be interpreted as follows: \gls{sc} tends to choose users with different \glspl{snr} as then the improvements over \gls{ts} are notable. Consequently, in the two-groups scenario, most likely one user from group ``A'' and one user from group ``B'' will be chosen. Thus, users in the least populated group are scheduled for transmission more frequently.
\item All users are drawing benefits from \gls{pf-sc} while this is not always the case for \gls{pf-ts}. In fact, the improvement in the throughput of group ``B'' is obtained by \gls{pf-ts} at the expense of the throughput of group ``A'' which decreases for large values of $\SNRav_\tr{B}$.
\end{enumerate}

Since in \gls{pf-sc} various users are simultaneously scheduled for transmission\footnote{We reuse the term ``scheduling'' to indicate that the power-fraction is not set to zero.} using \gls{sc}, it would be interesting to define how many can be simultaneously scheduled to allow the gains in \figref{Fig:Rate.A.B} to materialize. 

To this end, we denote by $K_\tr{A}$ and $K_\tr{B}$ the number of users scheduled for transmission in groups ``A'' and ``B'', respectively. In \figref{Fig:Prob.A.B}, we show the empirical probability of the events corresponding to different pairs $(K_\tr{A}, K_\tr{B})$ that are the most likely to occur and we observe that
\begin{enumerate}
\item In most cases, the number of users scheduled for simultaneous transmission is relatively small: \mbox{$\Pr\set{K_\tr{A}+K_\tr{B}\leq3}>0.7$}. It is an important observation as \gls{sc} with a small number of users might be realized via practical \gls{mcs} such as the standard-defined hierarchical modulation \cite{Chari07}. 
\item The event $K_\tr{A}+K_\tr{B}=1$ means that only one user is scheduled, which is likely to happen for $\SNRav_\tr{B}\approx\SNRav_\tr{A}$, \ie where \gls{sc} and \gls{ts} are equivalent. The probability of using \gls{sc} increases when $\SNRav_\tr{B}$ increases, \ie when the difference between \glspl{snr} becomes significant. 
\item The most likely to be scheduled are users taken from group ``B'' ($K_\tr{B}=2$ or  $K_\tr{B}=3$) but even then, one of the users from group ``A'' is also scheduled. This explains the gains of \gls{pf-sc}: while we privilege high-\gls{snr} users from group ``B''; we still feed data to low-\gls{snr} users from group ``A'' using \gls{sc}. 
\end{enumerate}

\begin{figure}[htb]
\psfrag{SNR-B}[ct][ct]{$\SNRav_\tr{B}$}
\psfrag{Proba}[c][c]{}
\psfrag{P10-XXX-XXX}[cl][cl][0.9]{$K_\tr{A}+K_\tr{B}=1$}
\psfrag{P11}[cl][cl][0.9]{$K_\tr{A}=1, K_\tr{B}=1$}
\psfrag{P12}[cl][cl][0.9]{$K_\tr{A}=1, K_\tr{B}=2$}
\psfrag{P13}[cl][cl][0.9]{$K_\tr{A}=1, K_\tr{B}=3$}
\psfrag{P21}[cl][cl][0.9]{$K_\tr{A}=2, K_\tr{B}=1$}
\begin{center}
\scalebox{1}{\includegraphics[width=0.7\linewidth]{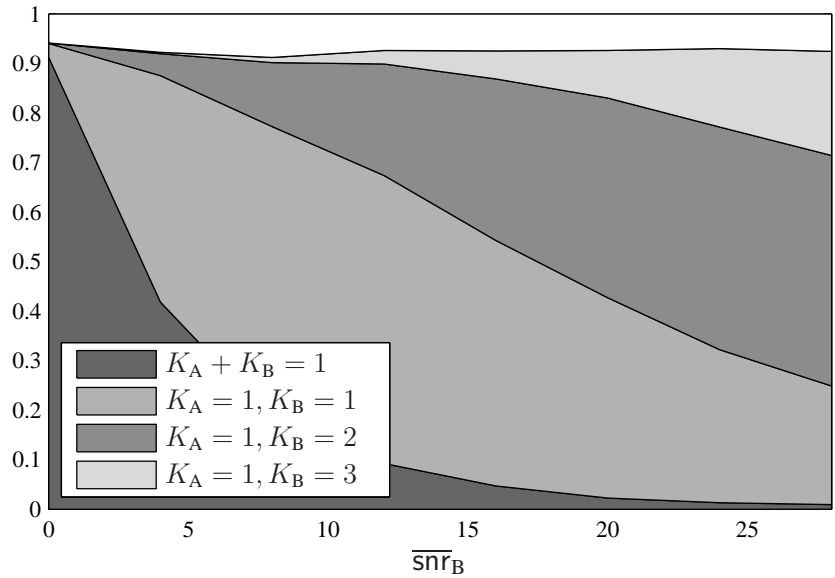}}\\
a)\\
\scalebox{1}{\includegraphics[width=0.7\linewidth]{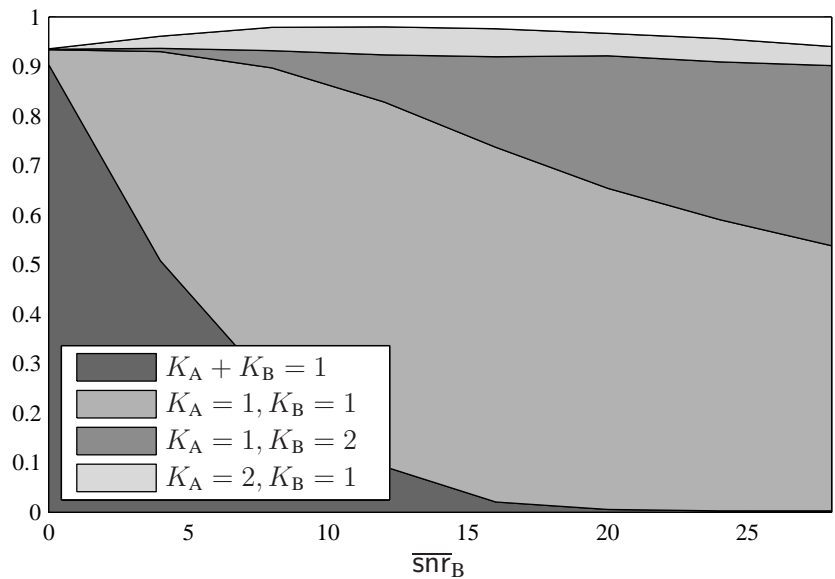}}\\
b)
\caption{The height of the shaded area corresponds to the probability of simultaneous transmission to $K_\tr{A}$ users in group ``A'' and $K_\tr{B}$ users in group ``B'' for a)~$L_\tr{A}=4$, $L_\tr{B}=16$ and b)~$L_\tr{A}=16$, $L_\tr{B}=4$; $\SNRav_\tr{A}=0$.}\label{Fig:Prob.A.B}
\end{center}
\end{figure}

We not that the number of scheduled users does not convey the whole information about the \gls{ra} outcome as it does not reflect the values of the power-fractions $p_{l}$ which, indeed, can be very  small. In particular, let us define
\begin{align}\label{}
P_{2}=\max_{j,k\in\mcL}\set{p_{j}+p_{k}}.
\end{align}
as the maximum power attributed to two users. We show in \figref{Fig:Prob.alpha.A.B} the empirical probability
$\Pr\set{ P_{2}\in \mcP}$, where $\mcP$ is the interval of power values. We can observe that, even if the probability of having more that two users scheduled for transmission in the scenario  $L_\tr{A}=4$, $L_\tr{B}=16$ is relatively large (\figref{Fig:Prob.A.B}a), the power assigned to additional users (beyond the first two users) is small. In fact, in 90\% of the analyzed cases, the first two users obtain more than 80\% of the power. We do not show the case  $L_\tr{A}=16$, $L_\tr{B}=4$ for which \mbox{$\Pr\set{ P_{2}\in (0.9,1]}>0.95$}, \ie almost all the available power is assigned to the first two users.

\begin{figure}[htb]
\psfrag{SNR-B}[ct][ct]{$\SNRav_\tr{B}$}
\psfrag{Proba}[c][c]{}
\psfrag{P0-70-XXX-XX}[cl][cl][0.9]{$\mcP=[0, 0.70)$}
\psfrag{P70-80}[cl][cl][0.9]{$\mcP=[0.70, 0.80)$}
\psfrag{P80-90}[cl][cl][0.9]{$\mcP=[0.80, 0.90)$}
\psfrag{P90-95}[cl][cl][0.9]{$\mcP=[0.90, 0.95)$}
\psfrag{P95-100}[cl][cl][0.9]{$\mcP=[0.95,1.0]$}
\begin{center}
\scalebox{1}{\includegraphics[width=1\linewidth]{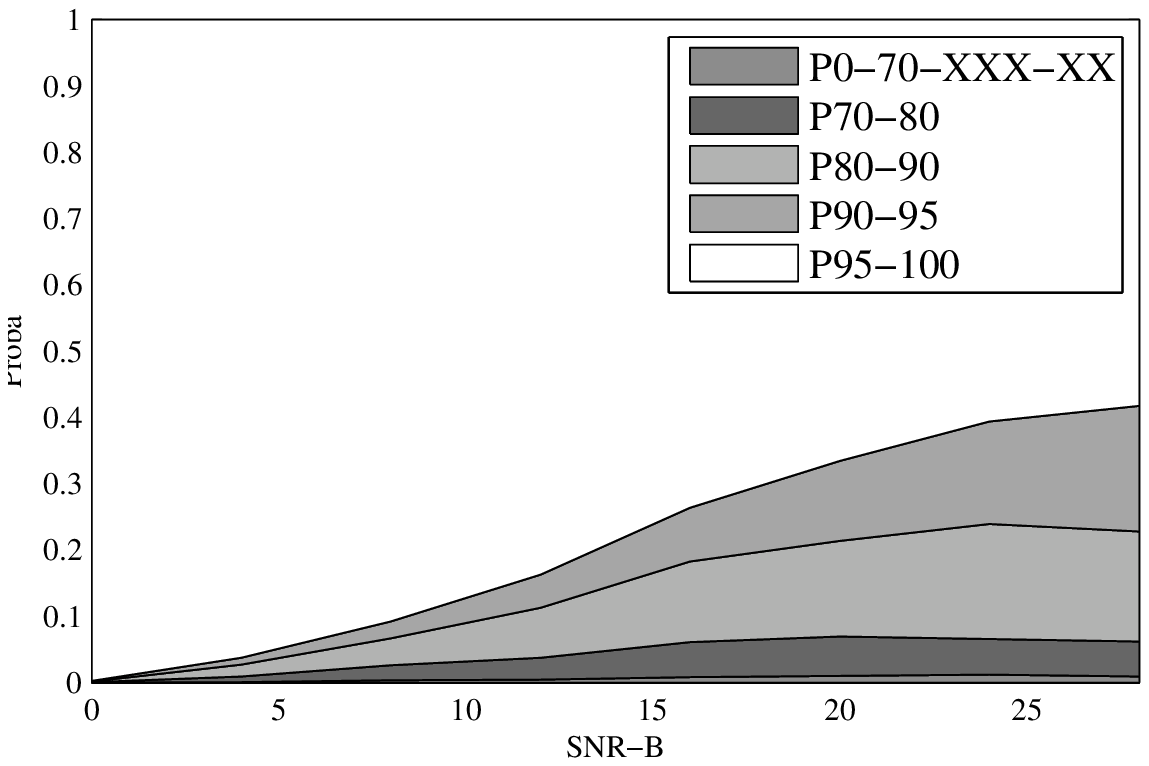}}
\caption{The height of the shaded area corresponds to the probability that the power of the two strongest users $P_{2}$ belongs to the inteval $\mcP$; $L_\tr{A}=4$ $L_\tr{B}=16$, and $\SNRav_\tr{A}=0$.}\label{Fig:Prob.alpha.A.B}
\end{center}
\end{figure}

\end{example}

\subsection{\gls{ra} under constraints on the number of scheduled users}\label{sec:RA.Kmax}

The numerical results in \exref{ex:AB.users} indicate that, with the optimal \gls{ra}, not only the number of users scheduled for transmission is small; but also the power of the first two users is dominant. This is interesting as, in practice, the number of superposed signals should not be very large. Thus, motivated by these results, we would like to obtain the \gls{ra} algorithm where we limit the number of users scheduled for transmission to a small value $K_\tr{max}$, and next, we will evaluate the penalty introduced by this additional constraint.

Our objective thus is to find the optimum indices $\hat{\mcL}=\set{\hat{\ell}_{1},\hat{\ell}_{2},\ld, \hat{\ell}_{K_\tr{max}}}$
\begin{align}\label{Opt.SC.K}
\hat{\mcL}&=\argmax_{\mcL\in \msL^{K_\tr{max}}} y_{\mcL}(\hat{\bp}_{\mcL}),
\end{align}
where
\begin{align}
\label{y.mcL.K}
y_{\mcL}(\hat{\bp}_{\mcL})&=\sum_{k=1}^{|\mcL|}\phi_{\ell_{k}}(\SNRvec,\hat{\bp}_{\mcL}),
\end{align}
and
\begin{align}
\label{Opt.SC.b.K}
\hat{\bp}_{\mcL}&=\argmax_{\bp} y_{\mcL}(\bp), \\ 
\nonumber
&\tnr{s.t.}\quad \sum_{k=1}^{|\mcL|}p_{\ell_{k}}=1, \quad p_{\ell_{k} }\geq 0 \\
\nonumber
&\qquad p_{l}=0\quad \tr{if}\quad l\notin\mcL
\end{align}
with $\msL^{K}$ being a $K$-fold Cartesian product of $\msL$.

This problem is more difficult than the optimization without constraint on the maximum number of allowed users $K_\tr{max}$. While the solution of \eqref{Opt.SC.b.K} has a linear complexity with $|\mc{L}|$, we have to repeat it for all the elements of the set $\msL^{K_\tr{max}}$; the overall complexity is then proportional to $L^{K_\tr{max}}$.

To avoid this polynomial complexity, we propose the greedy optimization algorithm described in Algorithm~\ref{algo.greedy}: starting with the empty set $\mcL=\emptyset$ we add one user at a time to maximize the overall objective function. While suboptimal, this algorithm provides a better solution than the \gls{ts}-based \gls{ra}. This is because the first user which is added to the set $\mcL$ is the one we find in the \gls{ts} approach, \cf \eqref{eq:pf.ts}. Other users are added to the set $\mcL$ solely if their presence improves the cost function. If this is not possible, and  the objective function does not increase (\ie the power-fraction attributed to the optimal user found in step \ref{step.opt} is zero $\hat{p}_{\hat{l}}=0$) the algorithm stops.

\renewcommand{\algorithmicrequire}{\textbf{Input:}}
\renewcommand{\algorithmicensure}{\textbf{Output:}}
\begin{algorithm}[htb]
{\small
\caption{Greedy maximization of the objective function: indices of active users are added to the set $\mcL$ one-by-one.}	\label{algo.greedy}
\begin{algorithmic}[1]
\REQUIRE $K_\tr{max}$
\ENSURE Suboptimal solution of the problem in \eqref{Opt.SC.K}.
\STATE $\mcL \gets \emptyset$
\STATE $K\gets 0$
\WHILE{$K\leq K_\tr{max}$} 
	\STATE $\hat{l}\gets \argmax_{l\in \msL, l \notin  \mcL} y_{\set{\mcL,l}}(\hat{\bp}_{\set{\mcL,l}})$\label{step.opt}
	\IF{$y_{\set{\mcL,\hat{l}}}\bigl(\hat{\bp}_{\set{\mcL,\hat{l}}}\bigr)>y_{\mcL}\bigl(\hat{\bp}_{\mcL}\bigr)$}
		\STATE $\mcL\gets \set{\mcL,\hat{l}}$
		\STATE $K\gets K+1$
	\ELSE
		\STATE \textbf{stop}
	\ENDIF
\ENDWHILE
\end{algorithmic}
}
\end{algorithm}

\begin{example}[Downlink transmission to users in a cell]\label{ex:cell}
Let us compare now \gls{pf-sc},  \gls{pf-ts},  and \gls{rr} resource allocation strategies in a scenario which will highlight the most important properties of the proposed \gls{ra} beyond the simplified case of two groups of users we considered in \exref{ex:AB.users}.


Consider the case when $L$ users are distributed over a circular cell with a normalized radius $d_\tr{max}=1$. 
We fix the \gls{snr} at the edge of the cell to $\SNRav(d_\tr{max})=0$dB and the average \gls{snr} at distance $d$ is given by $\SNRav(d)=d^{-\nu}$, where the path loss exponent is set to $\nu=3$ \cite{Liew08,Zafar13}. To avoid singularity (infinite \gls{snr}) at $d=0$, we set $\SNRav(d)=\SNRav(d_\tr{min})$ if $d\leq d_\tr{min}$ where $d_\tr{min}=0.1$, and the maximum average \gls{snr} is thus $\SNRav(d_\tr{min})=30$dB.

We assume that the users are uniformly distributed over  the cell and since only their distance $d$ to the \gls{bs}  is important, we generate the latter as $d=\sqrt{x}$, where $x$ is uniformly distributed in $(0,1)$. The positions of the users are randomly generated $N_\tr{rep}=1000$ times. Next, for all users whose distance falls into the interval \mbox{$[d- \Delta,d+\Delta]$}, we calculate the throughput averaged over $N_\tr{rep}$ realizations of users' positions. We denote it by $R(d)$ and show in \figref{Fig:Rate.Dist} for \gls{pf-ts}, \gls{pf-sc}, and \gls{rr} resource allocation strategies with $L=50$. 

These results are in line with the conclusions obtained from \exref{ex:AB.users}: the least populated groups of users (\ie those close to \gls{bs}) experience the greatest improvement in their throughput. For the case we analyze, when $d<0.35$ the increase is greater than 100\% and in the vicinity of the \gls{bs} we obtain a 300\% throughput gain.

At the same time, the throughput of all users is improved irrespectively of their distance $d$. This results in an increase of the aggregate throughput of the cell that we show in \figref{Fig:Rate.NUsers} as a function of the number of users $L$. We can appreciate that with respect to \gls{pf-ts}, the aggregate throughput of \gls{pf-sc} increases by 50\%  when $L>100$. 

As we have seen in \exref{ex:AB.users}, \gls{sc} tends to schedule more users with strong \gls{snr}, while keeping at least one weak-\gls{snr} user served. This explains the results of two-users \gls{sc} (denoted as $\SC_{2}$): the penalty due to  the constraint on the number of users $K_\tr{max}=2$ is more notable for strong-\gls{snr} users and is less important for users that are far from the \gls{bs}. Quite interestingly, there are no important differences between the throughput obtained via heuristic two-users \gls{ra} described in Algorithm~\ref{algo.greedy} and the optimal complex enumeration \eqref{Opt.SC.K}.

\begin{figure}[h]
\psfrag{distance}[ct][ct]{$d$}
\psfrag{ylabel}[c][c]{}
\psfrag{SC-XXX-XXX-XXX}[cl][cl][0.9]{$R(d), \SC$}
\psfrag{SC-U2}[cl][cl][0.9]{$R(d), \SC_2, \tr{Opt.}$}
\psfrag{SC-U2-H}[cl][cl][0.9]{$R(d), \SC_{2}, \tr{Greedy}$}
\psfrag{TS}[cl][cl][0.9]{$R(d), \TS$}
\psfrag{RR}[cl][cl][0.9]{$R(d), \RR$}
\begin{center}
\scalebox{1}{\includegraphics[width=1\linewidth]{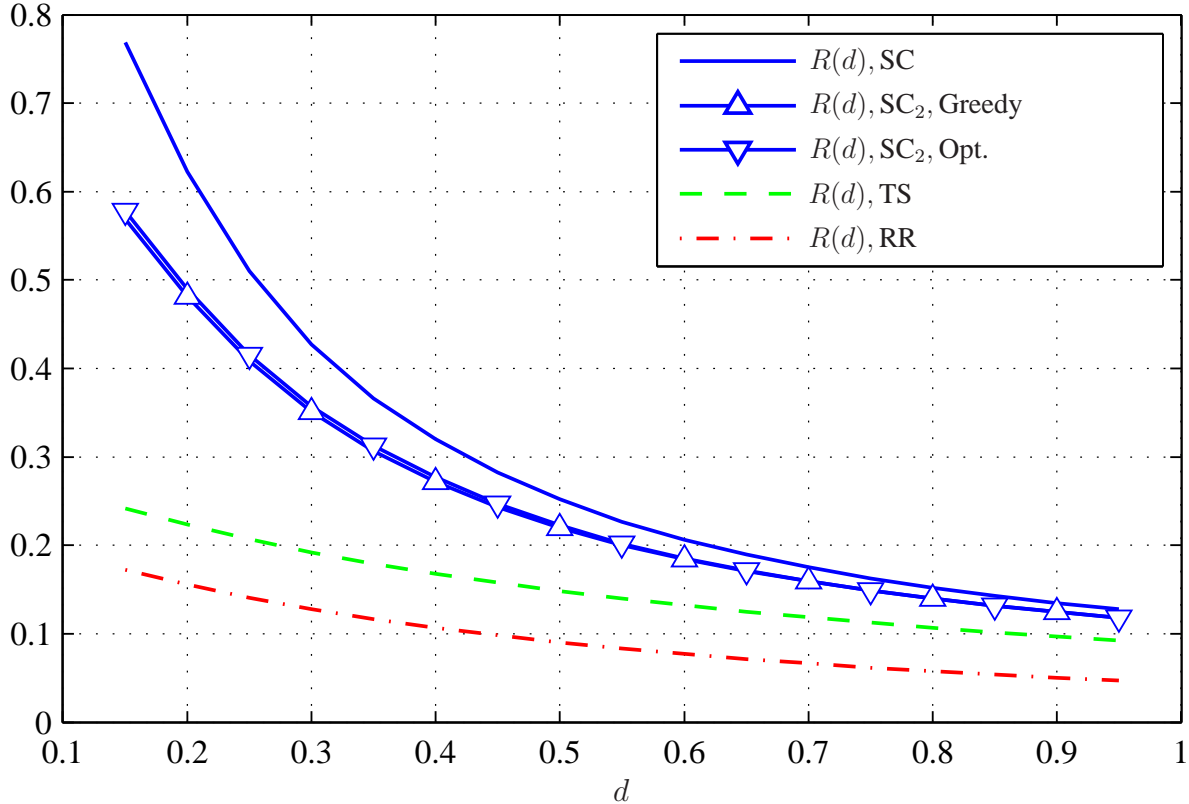}}\
\caption{The throughput as a function of the normalized distance $d$ of the user from the \gls{bs};  $L=50$, the average \gls{snr} at the cell's edge is given by $\SNRav(1)=0$dB; $\SC_{2}$ refers to \gls{sc} under constraint $K_\tr{max}=2$, ``Opt.'' to the optimal exhaustive search \eqref{Opt.SC.K}, and ``Greedy'' to the results obtained via Algorithm~\ref{algo.greedy}.}\label{Fig:Rate.Dist}
\end{center}
\end{figure}

\begin{figure}[h]
\psfrag{xlabel}[ct][ct]{$L$}
\psfrag{ylabel}[c][c]{$R$}
\psfrag{SC-XXX-XXX-X}[cl][cl][0.9]{$\SC$}
\psfrag{SC-U2}[cl][cl][0.9]{$\SC_{2}, \tr{Greedy}$}
\psfrag{TS}[cl][cl][0.9]{$\TS$}
\psfrag{RR}[cl][cl][0.9]{$\RR$}
\begin{center}
\scalebox{1}{\includegraphics[width=1\linewidth]{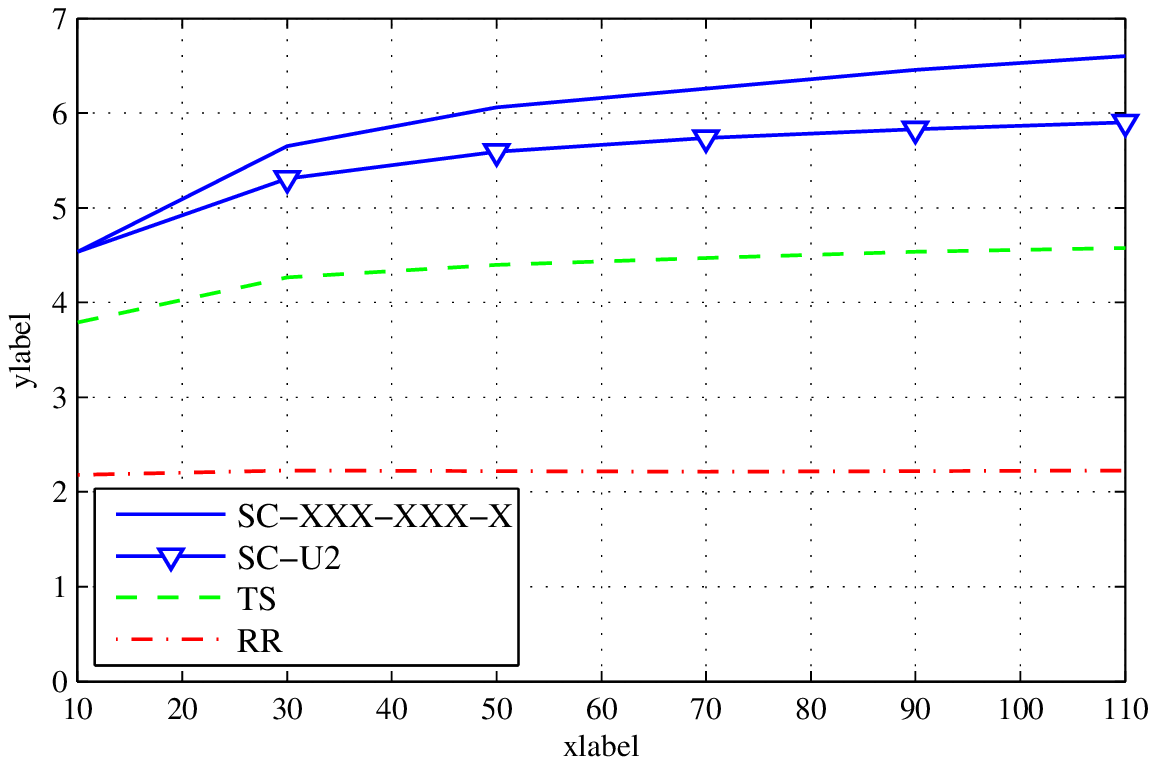}}\
\caption{The aggregate cell throughput as a function of the number of users $L$; the average \gls{snr} at the cell's edge is $\SNRav=0$dB.}\label{Fig:Rate.NUsers}
\end{center}
\end{figure}
\end{example}

\section{Conclusions}\label{Sec:Conclusions}
We analyzed the problem of transmitting  information to multiple users over a shared downlink wireless channel using \gls{sc}. We solved the problem of allocating the power to the users maximizing the criterion of sum of utility function and we have shown examples based of the criterion of proportional fairness. The proposed resource allocation algorithm easily deals with a very large number of users and we illustrated its operation with numerical examples showing a rate increase from 20\% and up to 300\%.
\section*{Appendix}\label{Sec:Appendix}

\begin{proof}[Proof of \propref{prop1}]
 It is convenient to  rewrite \eqref{SC_KKT_2} as
\begin{align}\label{SC_KKT_2.app}
\frac{\SNR_{j}\beta_{j}}{\SNR_{k}\beta_{k}}>f_{j,k}(\hat{\ov{p}}_{k}),
\end{align}
where the function 
\begin{align}\label{fjk.app}
f_{j,k}(p)\triangleq \frac{1+p\SNR_{j}}{1+p\SNR_{k}}
\end{align}
is  monotonically growing for $p\in(-1/\SNR_{k}, \infty)$. 

Therefore, to prove \eqref{Prop1.1} we have two cases to consider
\begin{enumerate}
\item For $\beta_{k}\leq\beta_{j}$ it is immediate to see that
\begin{align}\label{}
\frac{\SNR_{j}\beta_{j}}{\SNR_{k}\beta_{k}}\geq\frac{\SNR_{j}}{\SNR_{k}}=\lim_{p\rightarrow\infty}f_{j,k}(p)>\max_{p\in[0,1]}f_{j,k}(p)
\end{align}
so \eqref{SC_KKT_2.app} is satisfied for any $\hat{\ov{p}}_{k}\in[0,1]$ and thus $\hat{p}_{k}=0$. 	

\item For $\beta_{k}>\beta_{j}$, to satisfy \eqref{SC_KKT_2.app} irrespectively of $\hat{\ov{p}}_{k}$, we need the following
\begin{align}\label{}
\frac{\SNR_{j}\beta_{j}}{\SNR_{k}\beta_{k}}>\max_{p\in[0,1]}f_{j,k}(p)=\frac{1+\SNR_{j}}{1+\SNR_{k}}
\end{align}
which is equivalent to $\tau_{k}<\tau_{j}$.
\end{enumerate}

To prove \eqref{Prop1.3}, we note that if we satisfy
\begin{align}\label{SNR.tau}
\frac{\SNR_{j}\beta_{j}}{\SNR_{k}\beta_{k}}<\min_{p\in[0,1]}f_{j,k}(p)\leq f_{j,k}(\hat{\ov{p}}_{k}),
\end{align}
then $\displaystyle{\frac{\SNR_{j}\beta_{j}}{\SNR_{k}\beta_{k}}}\!<\!f_{j,k}(\hat{\ov{p}}_{k})$ is satisfied  irrespectively of $\hat{\ov{p}}_{k}$, and \eqref{SNR.tau} is equivalent to $\nu_{j}<\nu_{k}$.

This terminates the proof.
\end{proof}

\begin{proof}[Proof of  \propref{prop2}]

We establish first a simple relationship, namely, from \defref{def.1} we obtain $\frac{\SNR_{j}\beta_{j}}{\SNR_{k}\beta_{k}}=f_{j,k}(p_{j,k})$. Then,
\eqref{SC_KKT_2.app} is equivalent to $f_{j,k}(p_{j,k})>f_{j,k}(\hat{\ov{p}}_{k})$. Because of the monotonicity of $f_{j,k}(p)$, the latter  is also equivalent to the following conditions
\begin{align}
\label{Simple.1}
p_{j,k}&>\hat{\ov{p}}_{k} &\Rightarrow& &\hat{p}_{k}&=0\\
\label{Simple.2}
p_{j,k}&<\hat{\ov{p}}_{k} &\Rightarrow& &\hat{p}_{j}&=0,
\end{align} 
where we know that $\hat{\ov{p}}_{k}\in[0,1]$. 

To prove \propref{prop2}, we proceed  by contradiction: suppose that $0\leq p_{j,k}\leq p_{m,j}\leq 1$, $\sum^{j-1}_{l=k+1}\hat{p}_{l}=0$ and  $\sum^{m-1}_{l=j+1}\hat{p}_{l}=0$. But, we suppose that $\hat{p}_{j}>0$, and from \eqref{Simple.2} we obtain $p_{j,k}\geq\hat{\ov{p}}_{k}$, and from \eqref{Simple.1} we get $p_{m,j}\leq\hat{\ov{p}}_{j}$. Thus,
\begin{align}\label{pj.0}
\hat{\ov{p}}_{j}\geq p_{m,j} \geq p_{j,k}\geq\hat{\ov{p}}_{k}=\hat{\ov{p}}_{j}+\hat{p}_{j},
\end{align}
where the last equality follows from \eqref{ov.p} and $\sum^{j-1}_{l=k+1}\hat{p}_{l}=0$. To satisfy \eqref{pj.0}, we must set $\hat{p}_{j}=0$; which contradicts the assumption $\hat{p}_{j}>0$. 

This terminates the proof.
\end{proof}

\begin{proof}[Proof of \propref{prop3}]

We proceed by contradiction. Suppose there is a non-empty set $\mcJ=\set{j_{1},\ld,j_{K'}}$, which contains subsequent indices to the non-purged users with active positivity constraints, \ie  $\ell_{j_{l}}\in\mcL, l=1,\ld, K'$, and $\mu_{\ell_{j_{l}}}>0, l=1,\ld,K'$, and for any $k\notin \mcJ$ we must have $\ell_{k}<\ell_{j_{1}}$ or $\ell_{k}>\ell_{j_{K'}}$.

There are three possible cases then 
\begin{enumerate}
\item\label{case.2}
$j_{1}>1$, $j_{K'}=K$, and there is $k=j_{1}-1$ such that $\mu_{\ell_{k}}=0$.
\item \label{case.3}
$j_{1}=1$, $j_{K'}<K$, and there is $m=j_{K'}+1$ such that  $\mu_{\ell_{m}}=0$.
\item \label{case.1}
$j_{1}>1$, $j_{K'}<K'$, and there are $k=j_{1}-1$ and $m=j_{K'}+1$ such that $\mu_{\ell_{k}}=0$ and $\mu_{\ell_{m}}=0$.
\end{enumerate}

In case~\ref{case.2}), we know that
\begin{align}\label{SC_KKT_1}
\frac{\partial y_{\msL}(\hat{\bp})}{\partial p_{k}} &> \frac{\partial y_{\msL}(\hat{\bp})}{\partial p_{j_{1}}}\\
\label{j.k.0}
\frac{\SNR_{j_{1}}\beta_{j_{1}}}{\SNR_{k}\beta_{k}}&<\frac{1+\hat{\ov{p}}_{j_{1}}\SNR_{j_{1}}}{1+\hat{\ov{p}}_{j_{1}}\SNR_{k}}\\
\label{j.k.1}
\frac{\SNR_{j_{1}}\beta_{j_{1}}}{\SNR_{k}\beta_{k}}&<1,
\end{align}
where the transition from \eqref{j.k.0} to \eqref{j.k.1} is based on the fact that $j_{K'}=K$. Thus, $\hat{\ov{p}}_{j_{K'}}=\hat{\ov{p}}_{j_{1}}=0$. Since \eqref{j.k.1} is equivalent to $\nu_{j_{1}}<\nu_{k}$, this means that $j_{1}$ cannot be in the set $\mcL$ as it would be purged via Algorithm~\ref{Algo.2}. This is a contradiction, so case~\ref{case.2}) cannot occur.

In case~\ref{case.3}), we know that
\begin{align}\label{SC_KKT_1}
\frac{\partial y_{\msL}(\hat{\bp})}{\partial p_{m}} &> \frac{\partial y_{\msL}(\hat{\bp})}{\partial p_{j_{K'}}}\\
\label{j.k.0.b}
\frac{\SNR_{m}\beta_{m}}{\SNR_{j_{K'}}\beta_{j_{K'}}}&>\frac{1+\hat{\ov{p}}_{j_{K'}}\SNR_{m}}{1+\hat{\ov{p}}_{j_{K'}}\SNR_{j_{K'}}}\\
\label{j.k.1.b}
\frac{\SNR_{m}\beta_{m}}{\SNR_{j_{K'}}\beta_{j_{K'}}}&>\frac{1+\SNR_{m}}{1+\SNR_{j_{K'}}},
\end{align}
where the transition from \eqref{j.k.0.b} to \eqref{j.k.1.b} is based on the fact that $j_{1}=1$. Thus, $\hat{\ov{p}}_{j_{K'}}=\hat{\ov{p}}_{j_{1}}=1$. Since \eqref{j.k.1.b} is equivalent to $\tau_{j_{K'}}<\tau_{m}$, this means that $j_{K'}$ cannot be in the set $\mcL$ as it would be purged via Algorithm~\ref{Algo.1}. This is a contradiction so case~\ref{case.3}) cannot occur.

In case~\ref{case.1}), we know that
\begin{align}\label{SC_KKT_1}
\frac{\partial y_{\msL}(\hat{\bp})}{\partial p_{k}} &> \frac{\partial y_{\msL}(\hat{\bp})}{\partial p_{j_{1}}}\\
  \frac{\partial y_{\msL}(\hat{\bp})}{\partial p_{m}}& > \frac{\partial y_{\msL}(\hat{\bp})}{\partial p_{j_{K'}}},
\end{align}
therefore,
\begin{align}
\label{pp.1}
\frac{\SNR_{m}\beta_{m}}{\SNR_{j_{K'}}\beta_{j_{K'}}}&>\frac{1+\hat{\ov{p}}_{j_{K'}}\SNR_{m}}{1+\hat{\ov{p}}_{j_{K'}}\SNR_{j_{K'}}}\Rightarrow p_{\ell_{m},\ell_{j_{K'}}}>\hat{\ov{p}}_{j_{K'}}\\
\label{pp.2}
\frac{\SNR_{j_{1}}\beta_{j_{1}}}{\SNR_{k}\beta_{k}}&<\frac{1+\hat{\ov{p}}_{j_{1}}\SNR_{j_{1}}}{1+\hat{\ov{p}}_{j_{1}}\SNR_{k}}\Rightarrow p_{\ell_{j_{1}},\ell_{k}}<\hat{\ov{p}}_{j_{1}}
\end{align}
where \eqref{pp.1} is obtained from \eqref{Simple.1}, and \eqref{pp.2} is obtained from \eqref{Simple.2}. Since $\hat{\ov{p}}_{j_{1}}=\hat{\ov{p}}_{j_{K'}}$, combining \eqref{pp.1} and \eqref{pp.2} yields
\begin{align}\label{pp.3}
p_{\ell_{m},\ell_{m-1}}>p_{\ell_{k+1},\ell_{k}}.
\end{align}

Since the following relationship must hold after running Algorithm~\ref{Algo.3}
\begin{align}\label{pij.sorted}
p_{\ell_{2},\ell_{1}}>p_{\ell_{3},\ell_{2}}>\ld>p_{\ell_{K},\ell_{K-1}};
\end{align}
\eqref{pp.3} is in contradiction with \eqref{pij.sorted}, which means that case~\ref{case.1}) cannot occur.

Since none of possible cases can occur, we arrive at a contradiction with the assumption of having active constraints among non-purged users; this terminates the proof.
\end{proof}
\section*{Acknowledgement}
The authors would like to thank Prof. Long Le (INRS, Canada) for providing useful insight into the problem of resource allocation.

\balance


\end{document}